\documentclass[11pt,x11names,a4paper]{article}

\usepackage[utf8]{inputenc}
\usepackage{lmodern}
\usepackage[T1]{fontenc} 
\usepackage{microtype} 

\usepackage[a4paper, left=25mm, right=25mm, top=30mm, bottom=25mm]{geometry} 

\usepackage{abstract}

\usepackage{xcolor}

\usepackage{cite}
\usepackage{hyperref}
\hypersetup{
	colorlinks=true,
	linkcolor=Blue4,
	citecolor=Red4,
	urlcolor=Green4,
	linktoc=page
}

\usepackage{amsmath,amssymb,slashed,mathbbol,mathtools}
\numberwithin{equation}{section}

\newcommand{\rmi}{i}
\newcommand{\rme}{\mathrm{e}}
\newcommand{\rmd}{\mathrm{d}}

\newcommand{\be}{\begin{equation}}
\newcommand{\ee}{\end{equation}}
\newcommand{\dd}{\mathrm{d}}
\newcommand{\f}[2]{\frac{#1}{#2}}
\newcommand{\e}{\rme}
\newcommand{\Tr}{\text{Tr}\,}
\newcommand{\R}{\mathbf{R}}

\newcommand{\Z}{\mathbf{Z}}

\newcommand{\U}{\text{U}}
\newcommand{\SU}{\text{SU}}
\newcommand{\SO}{\text{SO}}

\newcommand{\SL}{\text{SL}}

\title{\fontsize{20pt}{24pt}\selectfont\textbf{One-loop quantization of Euclidean D3-branes in holographic backgrounds}\vspace{2mm}}

\author{
\large{\href{mailto:ffg@hi.is}{Fri{\dh}rik Freyr Gautason}$^1$ and \href{mailto:jvanmuid@sissa.it}{Jesse van Muiden}$^{2,3}$}\\[5mm]
{}$^1${\normalsize University of Iceland, Science Institute}\\
{\normalsize Dunhaga 3, 107 Reykjav{\'i}k, Iceland}\\[5mm]
{}$^2${\normalsize SISSA}\\
{\normalsize Via Bonomea 265, I 34136 Trieste, Italy}\\[5mm]
{}$^3${\normalsize INFN - Sezione di Trieste}\\
{\normalsize Via Valerio 2, I-34127 Trieste, Italy}
}

\date{}

\begin{document}
{\hypersetup{urlcolor=black}\maketitle}
\thispagestyle{empty}
\begin{abstract}
		In this note we analyze the semi-classical quantization of D3-branes in three different holographic backgrounds in type IIB string theory. The first background is Euclidean AdS$_5$ with $S^1\times S^3$ boundary accompanied with a twist to preserve supersymmetry. We work out the spectrum of fluctuations around the classical D3-brane configuration, compute its one-loop partition function, and match to the non-perturbative correction to the superconformal index of ${\cal N}=4$ SYM. We then study Euclidean D3-branes in the Pilch-Warner geometry dual to the IR Leigh-Strassler fixed point of ${\cal N}=1^*$ with the aim to find non-perturbative corrections to its index. Finally we study Euclidean D3-branes in the non-geometric ${\cal N}=2$ J-fold background which is dual to the gauging of the 3D Gaiotto-Witten SCFT. 
	\end{abstract}
	%
	%
	\section{Introduction}
	Since its discovery, the AdS/CFT correspondence has provided a wealth of knowledge in the context of both quantum field theory, and quantum gravity. Due to the advent of exact tools in quantum field theory, such as supersymmetric localization and the analytic bootstrap, the correspondence has been verified in a plethora of non-trivial ways. Additionally, it has proven to be a fruitful pathway to study string and M-theory away from their respective two-derivative low-energy supergravity regimes. The review  \cite{Gopakumar:2022kof} has  summarized recent progress and relevant references in this context.
	
	A particular venture that has sparked our interest is the semi-classical quantization of strings and branes in curved holographic backgrounds of type II- and M-theory. A full quantization of strings and branes in curved backgrounds has proven to be an arduous task. However, a combination of modern QFT tools,  and the AdS/CFT correspondence provides a systematic approach to achieve this goal in a semi-classical expansion, at the very least when considering supersymmetric systems. Indeed, recent examples where this approach has proven to be rewarding include the quantization of closed strings on disks with a hyperbolic metric, which are dual to expectation values of supersymmetric Wilson loops in the dual QFT, see for example \cite{Gautason:2021vfc,Giombi:2020mhz}. In a recent application we have also studied the quantization of strings on two-cycles of genus-0 with the round metric \cite{Gautason:2023igo}. In particular, we have shown in two cases that such a semi-classical quantization provides a precise holographic match with non-perturbative contributions to the dual QFT partition functions. One of the cases that was studied concerns strings in the type IIA holographic  background dual to five-dimensional SYM. This was obtained as a twisted dimensional reduction of  AdS$_7\times S^4$ \cite{Bobev:2018ugk}, which  is dual to the six-dimensional $(2,0)$ theory. In this background the semi-classical quantization of the multi-wound genus-0 strings reproduced the full tower of non-perturbative contributions to the $S^5$ partition function of the SYM theory in the large $N$ limit at strong coupling. This tower has a direct connection to the giant graviton expansion in the $(2,0)$ theory, as it is simply a Cardy-like limit of these giant gravitons in the AdS$_7\times S^4$ background of M-theory.\footnote{This result was partially generalized to include also finite $\beta$ effects to the thermal partition function by studying M2 branes in the aforementioned twisted 11d background in \cite{Beccaria:2023sph}.} Giant graviton expansions have received  interest recently as they point to the fact that the full quantum gravitational path integral  can in certain cases be computed by simply considering a single saddle (the thermal AdS background) and an infinite series of states consisting of branes on top of this saddle \cite{Arai:2019xmp,Imamura:2021ytr,Gaiotto:2021xce,Murthy:2022ien,Choi:2022ovw,Beccaria:2023hip,Lee:2023iil,Eleftheriou:2023jxr}.\footnote{This work has subsequently resulted in an exploration finding such structures in different theories across dimensions and number of preserved supersymmetries, some examples include \cite{Arai:2019aou,Arai:2020uwd}, and more recent work closely related to our approach taken in this paper can be found in \cite{Beccaria:2023sph,Beccaria:2023cuo}.} 
	\subsection{Indices and twists in holography}
	In this paper we study the semi-classical quantization of D3-branes in three distinct type IIB string theory backgrounds. These backgrounds are respectively dual to $\mathcal N=4$ SYM with gauge group $\text{SU}(N)$ on $S^3\times S^1$, the Leigh-Strassler CFT on $S^3\times S^1$, and finally the $\mathcal N=4$ SYM theory on $S^3\times S^1_{\mathfrak J}$, where the circle contains an S-duality wall. These QFTs can be localized in order to compute  their supersymmetric partition functions on $S^3\times S^1$ as a function of $N$. In the large $N$ limit it is expected that these partition functions can be factorized from the string theory point of view into a supergravity contribution, and towers of brane excitations wrapping Euclidean cycles in the internal geometry, reminiscent of the giant graviton expansions discussed above. For the $\mathcal N=4$ SYM and the Leigh-Strassler theory the partition functions are directly related to the Witten (or superconformal) indices through a multiplicative factor of the Casimir energy:
	\begin{equation}
		\mathcal Z_{S^1 \times  S^{3}} = \rme^{-\beta E_c} \mathcal I\,.
	\end{equation}
	For the S-fold theory the situation is more involved, and we will discuss this setup separately below. For the initial two cases, however, the object that one would like to compute is thus their Witten indices, which in an $\mathcal N=1$ language schematically can be expressed as
	\begin{equation}\label{Eq: general N=1 index}
		\mathcal I = \text{Tr}\,\rme^{2\pi\rmi (J+\bar J)} x^{\delta} \rme^{-\mu_i p_i}\,,\qquad \delta = H -2J - \frac32 r
	\end{equation}
	where $\delta$ is the anti-commutator of two supercharges, and the trace is taken over all states for which $\delta=0$, $J$ and $\bar J$ form the Cartan of the Lorentz group, $r$ is the superconformal $\text{U}(1)_R$ charge, and finally and $(\mu_i,p_i)$ are respectively the chemical potentials and charges of additional possible global symmetries in the theory. For the specific example of the $\mathcal N=4$ SYM theory, with gauge group $\text{SU}(N)$, this index reduces to
	\begin{equation}\label{Eq: 1/16 BPS index N=4}
		\mathcal I_{N} = \text{Tr}\,\rme^{2\pi\rmi (J+\bar J)}  x^{\delta} q^{H+J} y^{2\bar J} v^{R_1-R_2} w^{R_2-R_3} \,,\qquad \delta = H - 2J - (R_1 + R_2 + R_3)\,,
	\end{equation}
	where $R_i$ are the three Cartan generators of the R-symmetry group $\text{SO}(6)$. We will be particularly interested in two limits of this index. In the first limit, known as the 1/2-BPS limit, the index can be written in a simplified form compared to \eqref{Eq: 1/16 BPS index N=4}, by taking all fugacities equal to $1$ except $q$, and taking $x = \sqrt{q}$ such that
	\begin{equation}\label{Eq: 1/2-BPS index}
		\mathcal I_{N}^{\text{1/2-BPS}} = \text{Tr}\,\rme^{2\pi\rmi (J+\bar J)} q^{(3H-R_1 - R_2 -R_3)/2} \,,
	\end{equation}
	where the trace this time goes over all $1/2$-BPS states. This index can be written as follows \cite{Gaiotto:2021xce}\footnote{The Pochhammer symbol is defined as $(q)_N = \prod\limits_{n=1}^N (1-q^n)$.}
	\begin{equation}\label{halfBPSIndex}
		\mathcal I_{N}^{\text{1/2-BPS}} = \frac{1}{(q)_N} = \mathcal I_{\infty}^{\text{1/2-BPS}}\left( 1+ \sum\limits_{n=1}^\infty (-1)^n \frac{q^{n(n+1)/2}}{(q)_n} q^{n N}\right)\,,
	\end{equation}
	where $\mathcal I_{\infty}^{\text{1/2-BPS}} = 1/(q)_\infty$ arises in the large $N$ limit as the contributions of fluctuating  (super)-gravitons in AdS$_5$ \cite{Kinney:2005ej}. The expansion in $q^N$ that multiplies the large $N$ limit has been interpreted as the contribution of giant gravitons to the index. The term which scales as $q^{nN}$ can be thought of as the contribution of $n$ D3-branes in the saddle point expansion and its prefactor arises from the quantization of the D3-branes around their classical solution. 
	%
	%
	Recovering this series directly from quantized D3-branes has been the focus of the recent paper \cite{Eleftheriou:2023jxr}. We will briefly comment on their computation below. 
	
	The limit that we will be interested in, commonly referred to as the Schur limit \cite{Gadde:2011ik,Gadde:2011uv}, does not take the fugacity $y$ to be trivial, but instead scales it as $y = \sqrt{q}$. As the index is independent of $x$ we can choose it to equal $1$, such its full expression reduces to
	\begin{equation}
		\mathcal I_{N}^{\text{Schur}} = \text{Tr}\,\rme^{2\pi\rmi (J+\bar J)} q^{H+J + \bar J} \,,
	\end{equation} 
	where the trace is over states saturing two BPS bounds instead of one as in \eqref{Eq: 1/16 BPS index N=4} (see \cite{Arai:2019xmp} for explicit expressions of the BPS bounds). In this limit the index was computed analytically as a function of $N$ in \cite{Bourdier:2015wda} and takes the form
	\begin{equation}\label{Eq: full N=2 index}
		\mathcal I_N^{\text{Schur}} = \mathcal I_{\infty}^{\text{Schur}} \left(1+\sum\limits_{n=1}^\infty (-1)^n \left[
		\binom{N+n}{N}
		 +\binom{N+n-1}{N} \right] q^{n^2} q^{n N }\right)\,,
	\end{equation}
	where $\mathcal I_{\infty}^{\text{Schur}} = \text{PE}\left[\frac{q(2+q)}{(1-q)^2}\right]$ constitutes the contributions coming from the large $N$ limit. We are interested in the leading non-perturbative contribution (in $N$) in this expression which is
	\begin{equation}
		- N q^{N+1}\,.
	\end{equation}
	Note that, for both the $1/2$-BPS and the Schur index, the fugacity of the Hamiltonian equals $q$, which scales with the periodicity of the circle $q=\e^{-\beta}$. 
	
	The holographic dual of the $\mathcal N = 4$ SYM theory is the celebrated AdS$_5 \times S^5$ background of type IIB string theory. To study the aforementioned indices from the string theory point of view one has to impose $S^3 \times S^1$  boundary conditions on Euclidean AdS. Importantly, to ensure that the background remains supersymmetric when Euclidean time is periodically identified with period $\beta$, one must twist $S^1$ with an additional circle in the geometry. We will refer to this twisted Euclidean time $S^1_\beta$ as the thermal circle, and the corresponding AdS geometry is called thermal. The twist is necessary because the Killing spinors on AdS$_5\times S^5$ generically depend on all coordinates including Euclidean time. If Euclidean time is periodically identified without further modifications, the Killing spinors are not globally consistent and supersymmetry is broken. A solution to this problem is to perform a local coordinate transformation $\phi\to \phi + i\tau$, where $\tau$ is Euclidean time and $\phi$ parametrizes a $\U(1)$ direction in AdS$_5\times S^5$. This local coordinate transformation cannot be undone without breaking supersymmetry because of the global issues with the Killing spinors mentioned above. The specific choice of the $\U(1)$ that is twisted is in one-to-one correspondence with the fugacities turned on in the ${\cal N}=1$ index \eqref{Eq: general N=1 index}.  For the 1/2-BPS index the twist is done with the $\U(1)$ generated by the Cartan $R_1 + R_2 + R_3$, which in the geometry corresponds to the great circle of $S^5$ (we call this the \textit{1/2-BPS twist}). For the Schur index, instead, the twist involves the Cartan $J + \bar J$, which in the geometry corresponds to the great circle of $S^3$ inside AdS$_5$ (which we call the \textit{Schur twist}). A completely general twist may be implemented by forming a linear combination of all available Cartan generators and would be relevant for the study of the general ${\cal N}=1$ superconformal index. In this paper we will mostly focus on the Schur twist but also briefly mention the 1/2-BPS twist.
	A twist analogous to the latter was previously studied for the Euclidean supergravity  backgrounds dual to the six-dimensional (2,0) theory in \cite{Bobev:2018ugk,Gautason:2021vfc} (see also \cite{Beccaria:2023cuo}).
	
	The non-perturbative corrections to the superconformal index in the 1/2-BPS and Schur limit can be reproduced in the bulk from D3-branes wrapping Euclidean compact cycles in the geometry. More specifically, they wrap the thermal circle in AdS and an $S^3 \subset S^5$, making them explicitly sensitive to the supersymmetry preserving twists discussed above. We will explicitly show how to semi-classically quantize these D3-branes in order to reproduce the non-perturbative contributions to the Witten index from string theory. Our approach can be viewed complementary to the approach of Imamura et. al. who have localized the QFT living on the D3-branes to find the same result (see \cite{Arai:2020qaj} for the specific case of the Schur index), and functions as a proof of concept for the procedure that we subsequently also apply in the more intricate geometries that are holographically dual to the Leigh-Strassler, and the S-fold theories, both with a reduced amount of global- and supersymmetries.
	
	The Leigh-Strassler theory is the IR fixed point of an $\mathcal N=1$ RG flow originating in $\mathcal N=4$ SYM \cite{Leigh:1995ep}. Apart from $\mathcal N=1$ superconformal symmetry it also preserves an $\text{SU}(2)$ flavor symmetry. Its superconformal index can be written down in an equivalent form to \eqref{Eq: general N=1 index}, where one merely turns on a fugacity for the R-symmetry Cartan and the flavor symmetry. Since the  index is a protected quantity under continuous RG-flows one can essentially compute it in the weakly coupled UV regime a la R\"omelsberger \cite{Romelsberger:2007ec} (see also \cite{Gadde:2010en}). This procedure was explicitly worked out in the case of the Leigh-Strassler CFT in \cite{Bobev:2020lsk} to find that the superconformal index of the Leigh-Strassler theory equals
	\begin{equation}
		\mathcal I^{\text{LS}} = \text{Tr}\,\rme^{2\pi\rmi (J+\bar J)} x^\delta q^{H+J}y^{2\bar J} u^{R_F} \,,\qquad \delta = H -2 J - \frac34 (R_1 + R_2 + 2 R_3)\,,
	\end{equation}
	where $R_F$ denotes the cartan of the $\SU(2)$ flavor symmetry. In the large $N$ limit this index was matched to the full supergravity spectrum, including all KK-modes, in the dual string theory geometry \cite{Bobev:2020lsk}. We are interested in a limit of this index similar reminiscent of the $\mathcal N=4$ Schur index, namely where we scale $y=\sqrt{q}$ and trivialise all other fugacities. To the best of our knowledge an expression similar to that of the Schur index in $\mathcal N=4$ SYM in \eqref{Eq: full N=2 index} has not been found for the Leigh-Strassler theory, which is something we hope to address in future work. 
	
	In this paper we focus on the gravitational description of this theory, and in particular the non-perturbative D3-brane excitations in the type IIB string theory solution. Explicitly, the geometry is topologically of the form $\text{AdS}_5 \times S^5$, where the AdS metric contains a twist along the thermal circle, identical to the Schur twist in the  $\mathcal N=4$ SYM theory, and the five-sphere is squashed. Inspired by the $\mathcal N=4$ SYM index we conjecture that the non-perturbative contributions to the superconformal index of the Leigh-Strassler theory, at large $N$, are described by Euclidean D3-branes wrapping compact four-cycles in the geometry. We find the stable brane configurations which wrap a squashed three-sphere inside the five-sphere, and the thermal circle in AdS$_5$. We show that  the  D3-brane on-shell action is given by $S_{\text{cl}} = 3\beta N/2$, providing a prediction for the exponential suppression of the leading non-perturbative correction to the superconformal index of the dual QFT to scale as
	\begin{equation}
		\rme^{-3\beta N/2}\,.
	\end{equation}
	It will be  interesting to see if a large $N$ analysis in the dual QFT can reproduce this behavior in this setup. We also study the QFT living on the squashed D3-branes and determine its spectrum. This spectrum is a first step towards a full 1-loop analysis of the Euclidean D3-branes, which will in turn provide the prefactor to the exponentially suppressed contribution to the superconformal index. A computation which we plan to report on in the near future. It would also be interesting to compare to the giant graviton expansion in the Klebanov-Witten theory recently discussed in \cite{Fujiwara:2023azx}.
	
	The final example in which we study the quantization of D3-branes involves a two-parameter family of non-geometric backgrounds in type IIB string theory first described in \cite{Bobev:2023bxs}. Their background metrics take the following form
	\begin{equation}
		\text{AdS}_4 \times S^1_\mathfrak{J} \times S^5\,,
	\end{equation}
	where the five-sphere is squashed, and the $S^1$ contains an S-duality wall determined by an $\SL(2,\mathbf{Z})$ element that we denote with $\mathfrak{J}$. This family of backgrounds is dual to a conformal manifold of three-dimensional $\mathcal N=2$ strongly coupled CFTs with a $\U(1)$ flavor symmetry arising from compactifying $\mathcal N=4$ SYM on a circle with said S-duality wall. This conformal manifold contains two symmetry enhanced points, one where the flavor symmetry is enhanced to $\text{SU}(2)$, and one point where the supersymmetry is enhanced to $\mathcal N=4$. In the latter point the $S^3$ partition function of the CFT was shown to equal \cite{Assel:2018vtq,Terashima:2011qi,Ganor:2014pha}
	\begin{equation}
		\mathcal Z_N = \frac{q^{N^2/2}}{(q)_N}\,,\quad \text{with} \quad q = \rme^{-\beta}	\,,
	\end{equation}
	and $(q)_N$ denotes the Pochhammer symbol. Since the supersymmetric $S^3$ partition function is independent of exactly marginal couplings, this result holds all over the conformal manifold. Which was explicitly shown to hold in the large $N$ limit through a holographic analysis in \cite{Bobev:2021yya,Arav:2021gra,Guarino:2020gfe}. Furthermore, it was recently  shown in \cite{Bobev:2023bxs} that the full partition function of these theories can be split into a perturbative supergravity piece, and a tower of non-perturbative D-brane states on top of this leading saddle
	\begin{equation}\label{JfoldPF}
		\mathcal Z_N = \frac{q^{N^2/2}}{(q)_\infty} \left( 1 + \sum\limits_{n=1}^\infty (-1)^n \mathcal Z_n q^{n(N+1/2)} \right)\,,
	\end{equation}
	where we have explicitly matched the classical on-shell action to the exponential behavior of the non-perturbative contributions in this partition function. Upon interpreting the prefactor $q^{N^2/2}$ as the Casimir energy on $S^3 \times S^1$ this expression is much alike to the superconformal index of $\mathcal N=4$ SYM and thus one can pose the question if it is possible to define and explicitly compute a supersymmetric index where the thermal circle contains an S-duality wall, and if on the string theory side this index can be reproduced by a giant graviton like expansion. In this paper we take a first step towards a semi-classical quantization of D3-branes in this non-geometric backgrounds, similar as we have discussed for the thermal AdS backgrounds mentioned above, by computing the spectrum of fluctuations consisting of scalars, fermions, and a vector on the squashed D3-brane. To the best of our knowledge it is not known how to apply supersymmetric localization on such non-geometric curved backgrounds with S-duality walls,\footnote{As was applied for the compact D3-branes on a thermal circle embedded in geometric backgrounds, in for example \cite{Arai:2019xmp,Arai:2020qaj,Imamura:2021ytr}.} and thus it seems that the quantization of D3-branes is a fruitful approach to study such non-perturbative contributions in string theory. We plan to report on the full one-loop quantization of the D3-branes in this non-geometric background in the near future.

	This paper is structured as follows. In section \ref{Sec: D3s in thermal AdS} we study the semi-classical quantization of D3-branes in thermal AdS$_5 \times S^5$ and reproduce the leading non-perturbative correction to the Schur index of $\mathcal N=4$ SYM from string theory at 1-loop. We also comment briefly on the 1/2-BPS index of $\mathcal N=4$ SYM from this persepective. In Section \ref{Sec:LS} we study Euclidean D3-branes embedded in the thermal AdS backgrounds dual to the Leigh-Strassler CFT. Finally, in Section \ref{Sec: D3-branes in J-fold backgrounds} we study the semi-classical quantization of D3-branes in S-fold backgrounds of type IIB string theory. In appendix \ref{App: kappa symmetry} we provide the $\kappa$-symmetry analysis for the D3-branes we discuss in Section \ref{Sec: thermal AdS5}, showing that they preserve the expected amount of supercharges. \\[0.5cm]
	\noindent\textbf{Note added:} In the preparation of this manuscript \cite{Beccaria:2024vfx} appeared on the ArXiv which has significant overlap with Section \ref{Sec: D3s in thermal AdS} of our paper.
	\section{D3-branes in thermal AdS$_5\times S^5$}\label{Sec: D3s in thermal AdS}\label{Sec: thermal AdS5}
	The background we are interested in is thermal AdS$_5 \times S^5$ with supersymmetric $ S^1_\beta \times S^3$ boundary conditions. To preserve supersymmetry a globally non-trivial twist involving the Euclidean time must be imposed, as mentioned in the previous section. This is because the Killing spinors of AdS$_5\times S^5$ generally depends on all coordinates and so compactifying e.g. Euclidean time will eliminate the Killing spinors. One way to identify the correct choice of twisting needed for a particular observable is through a comparison to the dual field theory. In this case we are interested in the index given in \eqref{Eq: full N=2 index} which indicates that the twist needed is between the time-circle and the azimuthal angle of $S^3 \subset \text{AdS}_5$. This results in a ten-dimensional metric\footnote{
	We work in Einstein frame throughout this paper. This is relevant even for the metric \eqref{Eq: 10d metric} as appropriate factor of $g_s$ has been absorbed into the length scale $L$.
	For future reference we mention that we label the flat directions as $(\rho,\tau,\xi_1,\xi_2,\xi_3,\theta,\phi,\omega_1,\omega_2,\omega_3)$. }
	\begin{equation}\label{Eq: 10d metric}
		\rmd s_{10}^2 = L^2\Big(\rmd \rho^2 + \cosh^2\rho \,\rmd \tau ^2 + \sinh^2\rho \,\rmd \tilde\Omega_3^2 + \rmd \theta^2 + \cos^2 \theta\, \rmd \phi^2 + \sin^2\theta \,\rmd \Omega_3^2\Big)\,,
	\end{equation}
	where 
	\begin{equation}
	\begin{aligned}
		&\rmd \tilde\Omega_3^2 = \rmd \xi_1^2 + \sin^2\xi_1(\rmd \xi_2^2 + \sin^2\xi_2 (\rmd \xi_3 + \rmi  \rmd \tau)^2)\,,\\
		&\rmd \Omega_3^2 = \rmd \omega_1^2 + \sin^2\omega_1(\rmd \omega_2^2 + \sin^2\omega_2 \,\rmd \omega_3 ^2)\,.
	\end{aligned}
	\end{equation}
	The only additional non-trivial background field is the four-form gauge potential
	\begin{equation}
		C_4 = \rmi L^4 \sinh^4 \rho \,\rmd \tau \wedge \text{vol}_{\tilde S^3} + L^4\sin^4 \theta\, \rmd \phi  \wedge {\text{vol}}_{S^3}\,.
	\end{equation}
The length scale $L$ is related to the rank of the gauge group in the dual theory via
\be
L^4 = 4\pi N\ell_s^4 \,.
\ee

We are interested in studying non-perturbative corrections to the $S^1_\beta \times S^3$ partition function arising from D3-brane instantons. The embedding of the D3-brane in question is rather simple, it wraps the Euclidean time and the $S^3 \subset S^5$.\footnote{In a Lorenzian setting similar D3-branes were studied in \cite{Biswas:2006tj,Mandal:2006tk}, see also \cite{Mikhailov:2000ya} for an extended supersymmetry analysis of the embedding of such D3-brane giant gravitons.} To describe the dynamics on the world-volume of the brane we will work in static gauge throughout, such that its metric equals\footnote{We have stripped of the length scale $L$ from the D3-brane metric for simplification. The dependence on the length scale will be reintroduced at the level of the tension of the D3-brane.}
\begin{equation}
\rmd s_{\text{D3}}^2 = (\cosh^2\rho - \sinh^2 \rho\,\sin^2\xi_1\, \sin^2\xi_2  )\rmd \tau^2 + \sin^2\theta \,\rmd \Omega_3^2\,. 
\end{equation}
The embedding is fully specialized by minimizing the classical D3-brane action which in this case is equivalent to imposing that the extrinsic curvature is traceless
\begin{equation}\label{Eq: extrinsic curvature}
K^{ia}_{\phantom{ia}a} = 0 \,.
\end{equation}
Usually the right-hand-side of this equation receives a contribution from the coupling of the brane to background $F_5$, but in the current case this contribution vanishes.
This traceless requirement imposes the brane to localize at $(\rho,\theta,\xi_2) = (0,\pi/2,0)$,\footnote{There is another solution $(\theta,\xi_1,\xi_2) = (\pi/2,\pi/2,\pi/2)$ for which the D3-brane action takes the same value. A $\kappa$-symmetry analysis, which we present in Appendix \ref{App: kappa symmetry}, shows that this D3-brane is not supersymmetric.}
where its action and metric evaluate to
\begin{equation}\label{Ads5classicalSol}
S_{\text{cl}} =  N\beta\,,\qquad \rmd s_{\text{D3}}^2 = \rmd \tau^2 + \rmd \Omega_3^2\,.
\end{equation}
In Appendix \ref{App: kappa symmetry} we provide the associated $\kappa$-symmetry analysis of this brane, showing that it is $1/4$-BPS. The fact that the classical action of the D3-brane equals $N\beta$ is already a good sign as it means that the D3-partition function scales as
\be
{\cal Z}_\text{D3} \sim \e^{-\beta N} = q^N\,.
\ee
Our next task is to determine the prefactor which is obtained by performing a one-loop quantization of the fluctuations around the classical background. To this end we start by computing the spectrum of fluctuations in the next two subsection before computing the one-loop partition function itself.

	\subsection{Bosonic spectrum}\label{ssec:bosons}
Before studying the explicit spectrum of the bosonic fluctuations around the classical configuration \eqref{Ads5classicalSol}, we discuss some generalities of the bosonic D3-brane action.

We will denote the world-volume coordinates with $\zeta^a$, and the transverse fluctuations in tangent basis with $\Phi^i$. In this respect $a=1,\dots,4$ is a curved index on the brane world volume and $i=1,\dots,6$ is a flat index on the normal bundle.\footnote{For future reference we note that flat world-volume indices will be underlined.} The bosonic action of the D3-brane is given by
	\begin{equation}\label{bosonicaction}
	\begin{aligned}
		S^{B}_{\text{D3}} =& \frac{2\pi}{(2\pi \ell_s)^4 }\int \Big( \rmd^4 \zeta \sqrt{\text{det}(g + \e^{-\Phi/2}{\cal F})} + \rmi q_{\text{D3}} \, \big(C\wedge \e^{{\cal F}}\big)_4\Big)\,,\\
	\end{aligned}
	\end{equation}
	where $q_{\text{D3}}$ specifies the charge of the brane, ${\cal F} = f-B_2$, and $f$ denotes the worldvolume field-strength on the brane. We will focus on positively charged branes ($q_{\text{D3}} = 1$) as the only difference with a negatively charged brane is subtle with regard of the supercharges that are preserved, as we discuss in Appendix \ref{App: kappa symmetry}, which however is not of importance for the results we discuss here. This combination of fields also appears in the WZ coupling of the brane to the RR fields which in explicit terms takes the form
	\be
	\big(C\wedge \e^{{\cal F}}\big)_4 = C_4 + C_2\wedge {\cal F} + \f12 C_0\,{\cal F}\wedge {\cal F}\,.
	\ee
	For the D-brane configurations in this paper the background values of the field strength  and indeed the entire world-volume gauge field ${\cal F}$ vanishes. We will therefore often drop it in our expressions for simplicity. However, when considering the one-loop fluctuations the gauge field does play an important role.

	In static gauge the spectrum of the bosonic sector in the worldvolume theory is determined by the transverse fluctuation of the brane, denoted by $\Phi^i$, and the worldvolume field strength around their background values. Instead of explicitly expanding the D-brane action to second order we utilize standard geometric methods to split quantities up into to tangent bundle and normal bundle quantities (see \cite{Harvey:1999as,Forini:2015mca}). In particular, we find a natural gauge connection on the brane\footnote{Here we have dropped the worldvolume gauge field ${\cal F}$ as it vanishes for the backgrounds we consider.}
	\begin{equation}\label{Eq: pullbacks}
	\mathcal A_a^{\phantom{a}ij} = \Omega_a^{\phantom{a}ij} + {{\mathcal A}_F}_a^{\phantom{a}ij}\,,\qquad {{\mathcal A}_F}_{a}{}^{ij}= \frac{\rmi}{2\times 3!}\epsilon_{abcd}  F_5^{ijbcd}\,,
	\end{equation}
	where the first term is the pull-back of the ten-dimensional spin-connection and the second term is a result of the coupling of the D3-brane to the background RR-fields. Here two of the indices on the five-form are in the normal bundle while the other three are in the tangent bundle on which the world-volume Levi-Civita tensor acts on.
	The second order scalar Lagrangian arising from the expansion of the D3 action can then be written as\footnote{This expression holds with minor modification to the quadratic expansion of other D- and M-branes.}
	\begin{equation}\label{quadraticBosons}
		\mathcal L_s = \frac12 (D^i_{\phantom{i}j} \Phi^j)^2 - \frac12  (R^a_{\phantom{a}iaj}+K_{i}^{\phantom{i}ab}K_{j ab}  +\frac{i}{2\times 4!}\epsilon^{abcd}\nabla_i F_{jabcd}+{{\mathcal A}_F}_{aki}{{\mathcal A}_F}^{ak}_{\phantom{ak}j} )\Phi^i\Phi^j\,,
	\end{equation}
	where $R$ is the ten-dimensional Riemann curvature and the covariant derivative includes, apart from the world-volume connection, also the gauge connection in \eqref{Eq: pullbacks}
	\begin{equation}
	D^{ij} = \delta^{ij}\nabla + \mathcal A^{ij}\,.
	\end{equation}

	Coming back now to the specific case of D3-branes in thermal AdS$_5$, evaluating the terms in this action and diagonalizing the gauge connection we find we can rewrite the full second order scalar Lagrangian simply as a set of six conformally coupled scalars
	\begin{equation}
		\mathcal L_s = \frac12\Big[(D\Phi)^2 + \frac1{6} \mathcal R \Phi^2\Big]\,, 
	\end{equation}
	with $\mathcal R = 6$ is the scalar curvature of the D3-brane and the gauge covariant derivative is defined as
	\begin{equation}
		D = \nabla_{S^1_\beta \times S^3} - i Q \mathcal A\,,\qquad {\mathcal A}= i\dd\tau\,.
	\end{equation}
The fact that the system reduces to such a simple action for the scalars is perhaps not surprising since we are just dealing with D3-branes on a conformally flat background. The non-trivial aspect of this theory is entirely contained in the constant imaginary gauge potential and the couplings of the fields to it.
Here we have introduced the charges $Q$ of each of the six modes and we find
\be\label{qforscalars}
Q\in\{0,0,1,-1,2,-2\}\,.
\ee
	
	To analyze this further we want to perform a KK reduction along $S^1_\beta$. To this end it is convenient to define the following complex scalar degrees of freedom
	\begin{equation}
		x = \frac{1}{\sqrt{2}} (\Phi_1 + \rmi \Phi_2)\,,\quad y = \frac{1}{\sqrt{2}} (\Phi_3 + \rmi \Phi_4)\,,\quad z = \frac{1}{\sqrt{2}} (\Phi_5 + \rmi \Phi_6)\,.
	\end{equation}
	Upon integration by parts, the full scalar Lagrangian can be expanded to find
	\begin{equation}
	\begin{aligned}
		\mathcal L_s =&\bar x \mathcal K_{0} x - {\bar x}\ddot x   + \bar y  \mathcal K_{0} y - {\bar y}(y-2 \dot y+\ddot y) + \bar z  \mathcal K_{0} z - {\bar z}(4  z- 4 \dot z+\ddot z ) \,,
	\end{aligned}
	\end{equation}
	where we have conveniently combined the mass $1$ with the 3D spherical Laplacian which is appropriate for conformally coupled scalars
	\begin{equation}
		\mathcal K_{0} = -\nabla^2_{S^3} + 1\,.
	\end{equation}
	The dot denotes the $\tau$-derivative which we deal with by imposing a standard KK-ansatz along $S^1_\beta$ which reduces to the problem to three towers of complex  3D scalar fields, with masses 
	\begin{equation}
		\quad M_n^2 = \left(n_\beta + \rmi Q\right)^2\,,\quad Q\in\{0,1,2\}
	\end{equation}	
	where $n_\beta = 2\pi n/\beta$, and $n\in \mathbf{Z}$ denotes the KK-level of the scalar. 
	
	Recall that the gauge field vanishes on-shell and so its fluctuations are easily evaluated by expanding in $f$ to second order. After rescaling the gauge field to absorb $g_s$, this results in the standard Maxwell term 
	\begin{equation}\label{Eq: Maxwell term}
		\mathcal L_A = \frac14 f^{ab}f_{ab}\,.
	\end{equation}

\subsection{Fermionic spectrum}\label{ssec:fermions}
We start from the fermionic action in \cite{Marolf:2003ye,Marolf:2003vf,Martucci:2005rb}.\footnote{We use the same conventions as in \cite{Martucci:2005rb} except we work in Einstein frame and we have reversed the sign of $B_2$.} After setting the worldvolume gauge field ${\cal F}$ to zero and pulling the $L^2$ out of the worldvolume metric the action takes the form
\be\label{prekappafermions}
S_F = \frac{iL^4}{(2\pi\ell_s)^4 }\int \dd^4\zeta \sqrt{\gamma} \bar\Theta\frac{1-\Gamma_\text{D3}}{2}\e^{-\Phi/4}\big[\Gamma^a\hat D_a-\hat\Delta\big]\Theta\,,
\ee
where $\Gamma_a = \partial_a X^M\Gamma_M$ are the gamma matrices pulled back to the world volume, $\gamma_{ab}$ is the metric on the D3, and $\Theta$ is a 32 component spinor in ten dimensions written as a pair of chiral fermions $\Theta= (\theta_1,\theta_2)$, for which $\Gamma_{11}\theta_n = \theta_n$. Finally, we have also defined
\be
\begin{split}
\hat D_a &= \partial_a +\frac14 \partial_a X^M\Omega_M{}^{AB}\Gamma_{AB} - \f1{4}\e^{-\Phi/2}\slashed{H}_{a}\sigma_3\\
&\qquad+\frac{1}{8}\Big[\e^{\Phi}\slashed{F}_1+\e^{\Phi/2}\slashed{F}_3\sigma_3+\f12\slashed{F}_5 \Big](\rmi\sigma_2)\Gamma_a\,,\\
\hat\Delta &= \f12\Big[ \slashed{\partial}\Phi -\f12\e^{-\Phi/2}\slashed{H}\sigma_3 -\e^{\Phi}\slashed{F}_1 (\rmi \sigma_2) -\f12
\e^{\Phi/2}\slashed{F}_3\sigma_1\Big]\,,
\end{split}
\ee
where the pauli-matrices act on the pair of chiral spinors and $\slashed{H}_{a} = \slashed{H}_{a\mu\nu}\Gamma^{\mu\nu}/2$. 
We also note that $\Gamma_\text{D3}$ is a projection operator on the brane which for vanishing worldvolume flux takes the simple expression\footnote{This expression differs slightly from \cite{Martucci:2005rb} since we are working with Euclidean D3-branes.}
\be
\Gamma_\text{D3}=\Gamma_{(4)}\sigma_2\,,
\ee
where $\Gamma_{(4)}$ is the chirality operator on the four-dimensional brane world-volume, i.e. the product of all pull-backed $\Gamma$-matrices with flat indices.
A common way to fix the kappa symmetry gauge in type IIB is to impose $\theta_1 =\theta_2 \equiv \theta$. This reduces the fermionic action to
\be
\begin{split}
S_F &= \frac{L^4i}{(2\pi\ell_s)^4  }\int \dd^4\zeta \sqrt{\gamma} \bar\theta \e^{-\Phi/4}\Big[\Gamma^aD_a - \Delta\Big]\theta\,,\\
D_a &= \partial_a +\frac14 \partial_a X^M\omega_M{}^{AB}\Gamma_{AB} - \f\rmi{4}\e^{-\Phi/2}\Gamma_{(4)}\slashed{H}_{a}+\frac{1}{8}\Big[\rmi\e^{\Phi}\Gamma_{(4)}\slashed{F}_1+\e^{\Phi/2}\slashed{F}_3+\f\rmi2\Gamma_{(4)}\slashed{F}_5 \Big]\Gamma_a\,,\\
\Delta &= \f12\Big[ \slashed{\partial}\Phi +\f\rmi2\e^{-\Phi/2}\Gamma_{(4)}\slashed{H} +\rmi\e^{\Phi}\Gamma_{(4)}\slashed{F}_1  
-\f12\e^{\Phi/2}\slashed{F}_3\Big]\,.
\end{split}
\ee
In the next sections we will choose a different kappa symmetry gauge which is more appropriate there. For the current case of interest, however, the standard gauge is applicable as most of the terms drop out. Note that when the dilaton is constant, the fermions should be rescaled by appropriate powers of $g_s$ to absorb the explicit factor of $\e^{-\Phi/4}$ in the action.

After inserting the background for twisted AdS$_5\times S^5$, the first term in the action gives rise to the Dirac operator on $S^1_\beta\times S^3$ plus a constant gauge connection along the $\tau$-direction. The last term depends on the RR field strength $F_5$, which when slashed takes the form
	\begin{equation}
		\slashed{F}_5 = \f{4}{L}( \rmi \Gamma_{12345}+\Gamma_{678910} )\,.
	\end{equation}
This also gives a coupling to the same constant gauge connection ${\mathcal A} = i \dd \tau$. Unsurprisingly, this turns out to be the same constant gauge potential that was felt by the scalar fields. The resulting operator
\be
i{\mathcal D} =i\slashed{\nabla}_{S^1_\beta\times S^3} + Q\slashed{\mathcal A}\,,
\ee
is the kinetic operator of four massless fermions on $S^1_\beta\times S^3$ which are charged with respect to the background gauge field as controlled by the charges of the fermions 
\be
Q = \frac{ 2s_1+s_2}{2}\in\left\{ -\frac12,\frac12,-\frac32,\frac32 \right\}\,,
\ee
which are obtained by taking all possible combinations of $s_{1,2}=\pm1$. The signs $s_{1,2}$ arise when reducing the ten-dimensional fermions to four dimensions, and can be thought of as the eigenvalues of the original spinors with respect to the ten-dimensional gamma matrices $\rmi\Gamma_{45}$ and $\rmi\Gamma_{67}$.

We now continue by imposing a standard KK-ansatz on the thermal circle and find that the fermionic operator takes the form
	\begin{equation}
i{\mathcal D} =i\slashed{\nabla}_{ S^3} + M_n\,,\qquad M_n=\pm(n_\beta + \rmi Q )
	\end{equation}
	where once again $n_\beta$ denotes the KK-level on the time circle and the sign $\pm$ in the mass is a result of the gamma-matrix along the $\tau$-direction in the original four-dimensional theory which originally multiplies the $M_n$-term. To properly write this as a three-dimensional operator, we should reduce the four-dimensional spinor to three-dimensions using $\Gamma_\tau$ to split them up. This means that half of the three-dimensional spinors will have mass $n_\beta+\rmi Q$ and the other half will have negative that mass. In the end this sign does not affect the one-loop partition function of the modes and so effectively the degeneracy of the spinors is simply doubled. We will therefore drop this sign in what follows.
	Before moving on to the evaluation of the one-loop determinants we list the full spectrum in Table \ref{Tab: spectrum}.
	\begin{table}[h]
	\centering
	\begin{tabular}{lcccc}\hline\hline
	 4D Field& Degeneracy & $|Q|$ &  $M^2$ & KK mass  \\\hline
	 Scalars& $2$ & $0$   & $1$ & $n_\beta  $\\
	 & $2$ & $1$  & $1$ & $(n_\beta + \rmi  )$\\
	 & $2$ & $2$  & $1$ & $(n_\beta + 2\rmi  )$\\\hline
	 Fermions& $2(4)$ &  $3/2$  & $0$ & $(n_\beta + 3\rmi/2 )$\\
	 & $2(4)$ &  $1/2$  & $0$ & $(n_\beta + \rmi/2 )$\\  \hline
	 Vector& $1$ &  $0$  & $0$ &$n_\beta $\\
	\end{tabular}
	\caption{A summary of the spectrum of  conformally coupled scalars, fermions, and vector. The fermions get doubled when reducing to three dimensions which is indicated by the doubled degeneracy in a bracket.}\label{Tab: spectrum} 
	\end{table}
	\subsection{One-loop determinants}
	Quantization of the above fluctuations boils down to evaluating the determinants of the operators derived above. To this end we must first recall the spectrum of the Dirac and Laplace operators on $S^3$ and multiply all eigenvalues together accounting for all modes. These products usually diverge and in order to regularize each individual contributions to the log of the determinants coming from the scalars, vectors and fermions we apply the regularization scheme used in \cite{Giombi:2014yra} (see also \cite{Beccaria:2023cuo}). 

	For all our fields the quadratic one-loop operator takes the form
	\be
	{\cal K} = {\cal K}_0  + M_n^2\,,
	\ee
	where ${\cal K}_0$ is the Laplacian on $S^3$ for the appropriate field with the conformal coupling taken into account.
	Let $\omega_k^2$ be the spectrum of ${\cal K}_0$ with degeneracy $d_k$. The logarithm of the determinant of ${\cal K}$ can be expressed as a double sum
	\begin{equation}\label{Eq: general form 1-loop}
	\begin{aligned}
		\f12\Tr\log{\cal K} =& \sum\limits_{n\in \mathbf{Z}} \sum\limits_{k=0}^\infty d_k \log[\omega_k^2 + M_n^2]\\
		 =&  \frac12\sum\limits_{n\in \mathbf{Z}} \sum\limits_{k=0}^\infty d_k (\log [(\omega_k^+)^2 + n_\beta^2] + \log [(\omega_k^-)^2 + n_\beta^2]) \,,
	\end{aligned}
	\end{equation}
	where we have used that the sum over $n$ runs over positive and negative integers and we have defined
	\begin{equation}
		M_n = n_\beta+\rmi Q \,,\quad \text{and}\quad \omega_k^\pm = \omega_k \pm Q\,.	
	\end{equation}
	The double sums in \eqref{Eq: general form 1-loop} are regularized as\cite{Giombi:2014yra}
	\begin{equation}
		\f12\Tr\log{\cal K} = \f{\beta}{2} \sum\limits_{k\in\mathbf{N}} d_k \omega_k + \f12\sum\limits_{k\in\mathbf{N}} d_k \Big(\log[1-\rme^{-\beta \omega_k^+}]+ \log[1-\rme^{-\beta \omega_k^-}]\Big) \equiv \beta E_c + \bar F(\beta,Q)\,,
	\end{equation}
	with $E_c$ the Casimir energy. Exponentiation gives the 1-loop contribution to the D3-brane partition function
	\begin{equation}
		\det{\cal K} = \rme^{-\beta E_c} \rme^{-\bar F(\beta,Q)} = \rme^{-\beta E_c} \text{PE}[\hat Z(\beta,Q)]\,,
	\end{equation}
	where PE represents the plethystic exponential, and $\hat Z(\beta)$ is the single particle index
	\begin{equation}
		\hat Z(\beta,Q) = \f12\sum\limits_{k\in\mathbf{N}}d_k (q^{\omega^+_k}+q^{\omega^-_k})\,,\quad \text{where} \quad q=\rme^{-\beta}\,.
	\end{equation}
	Computing the single particle index can often be much simpler than computing the full partition function and is a useful intermediate step. It is also useful because it clearly identifies the zero mode contributions that one has to treat separately. 
	
	The spectrum of Laplace operators for conformally coupled scalars, fermions and vector  on $S^3$ is well known and can be summarized by\footnote{The spectrum of the kinetic operators on spheres for scalars, fermions and vectors can be found in e.g. \cite{Kutasov:2000td}.}
	\begin{equation}
	\begin{aligned}
	&\text{scalars}:\quad	 	&\omega_k^2 = (k+1)^2\,, \hspace{1cm} &d_k = (k+1)^2\,,\\
	&\text{fermions}:\quad	 	&\omega_k^2 = (k+\frac32)^2\,,\hspace{1cm} &d_k  = \frac12 (k+1)(k+2)\,,\\
	&\text{vectors}:\quad	 	&\omega_k^2 = (k+2)^2\,,\hspace{1cm} &d_k  = 2(k+1)(k+3)\,,
	\end{aligned}
	\end{equation}
	where $k=0,1,\ldots$ Before evaluating the full 1-loop partition function we note that the spectrum contains $10$ scalar zero modes, and $8$ fermionic zero modes, at KK-level $n=0$
	\begin{equation}\label{Eq: zero modes}
	\begin{aligned}
		&2\times \text{scalars}:\quad \hspace{0.32cm} k=0\,,\quad d_k = 1,\quad Q^2 = 1 \hspace{1cm} \rightarrow \quad 2 \times \text{zero-modes}\,,\\
		&2\times \text{scalars}:\quad \hspace{0.32cm} k=1\,,\quad d_k = 4,\quad Q^2 = 4 \hspace{1cm} \rightarrow \quad 8 \times \text{zero-modes}\,,\\
		&2\times \text{fermions}:\quad k=0\,,\quad d_k = 1,\quad Q^2 = 9/4 \hspace{0.6cm} \rightarrow \quad 8 \times \text{zero-modes}\,,
	\end{aligned}
	\end{equation}
	which we will treat separately after computing the single letter index of the fields. Applying the regularization procedure above, utilizing the mass spectrum summarized in Table \ref{Tab: spectrum}, we find the following Casimir energies and single letter indices for the scalars, fermions, and the vector
	\begin{equation}
		\begin{aligned}
			E_c^{\Phi} =& \sum\limits_{Q = 0}^{2} E_c(Q) = \frac{31}{40}\,, \quad E_c^{\Psi} = \,2\sum\limits_{Q = \frac{-3}{2}}^{\frac32} E_c(Q) = \frac{58}{15}\,,\quad E_c^{A} = \frac{251}{120}\,,
		\end{aligned}
	\end{equation}
	and 
	\begin{equation}
		\begin{aligned}
		\hat Z^{\Phi}(\beta) =& \sum\limits_{Q = 0}^2 \hat Z^{\Phi}(\beta,Q)= \frac{(1+q)(1+q^2)(1+q+q^2)}{(1-q)^3 q}\,,\\
		\hat Z^{\Psi}(\beta) =& \,4\sum\limits_{Q = \frac{-3}{2}}^{\frac32} \hat Z^{\Psi}(\beta,Q) = \frac{4 (1+q)(1+q^2)}{(1-q)^3}\,,\\
		\hat Z^{A}(\beta) =& \frac{2 q^2(3-q)}{(1-q)^3}\,.
	\end{aligned}
	\end{equation}
	Adding the contributions together we find the following simple result
	\begin{equation}
	\begin{aligned}
		E_c =& E_c^\Phi - E_c^\Psi + E_c^A = -1\,,\\
		\hat Z_1 =& \hat Z^\phi - \hat Z^\Psi + \hat Z^A = 1 + \frac1q - q\,.
	\end{aligned}
	\end{equation}
	The constant contribution $1$ in the single particle index arises due to the mismatch in scalar and fermion zero-modes listed in \eqref{Eq: zero modes}. These, as well as the paired zero-modes should be separated off when computing the partition function and  we therefore subtract the constant contribution  from the single particle index before taking its plethystic exponential. The Plethystic exponential of the remaining single particle index can formally be evaluated to equal $-q$ (see e.g. \cite{Imamura:2021ytr}), such that the full one-loop partition function of the D3-brane is 
	\begin{equation}
		\mathcal Z_\text{D3}(\beta) = Z_{\text{zero-modes}}\rme^{-\beta E_c} \,\text{PE}[\hat Z_1(\beta)] \e^{-S_\text{cl}}= -Z_{\text{zero-modes}}\rme^{-\beta E_c} q^{1+N}\,.
	\end{equation}
	As mentioned, the zero-modes can be split into eight pairs of fermion and boson zero-modes, and two unpaired bosonic zero-modes. In \cite{Gautason:2023igo} we argued that the paired zero-modes on a quantized string of genus-0 contributed a factor two in the 1-loop partition function arising from the fact that the string can localize at two points in the target space geometry. This was justified by deforming the background geometry, essentially giving rise to a mass for the zero-modes which localized them. The factor two could therefore be a posteriori justified by carefully counting the number of points where $\U(1)$ isometries degenerate. Using similar counting arguments in the current situation is slightly problematic since the location of the D3-brane is fixed to a single position at the classical level, in light of this we postulate the contribution of the paired zero-modes to be one. Finally, one has to integrate over the collective coordinates associated to the two remaining scalar zero-modes, which we can identify with the spherical coordinates $\xi_1$ and $\xi_3$ to find that
	%
	\begin{equation}
	Z_{\text{zero-modes}} 
	\sim T_{D3} 2 \pi^2 = N\,.
	\end{equation}
	The final answer for the 1-loop partition function of the D3-branes is thus
	\begin{equation}
		\mathcal Z_\text{D3}(\beta) = -\rme^{-\beta E_c} N q^{1+N}\,,
	\end{equation}
matching with the QFT result.

\subsection{Comment on the 1/2-BPS index}\label{Sec: 1/2 bps index}
We close this section by analyzing the 1/2-BPS index using semi-classical D3-branes. As discussed in the introduction, depending on the index under consideration, the twist changes. For the particular case of 1/2-BPS index the twist involves the equator of $S^5$ as follows
\begin{equation}\label{halfBPSmetric}
		\rmd s_{10}^2 = L^2\Big(\rmd \rho^2 + \cosh^2\rho \,\rmd \tau ^2 + \sinh^2\rho \,\rmd \tilde \Omega_3^2 + \rmd \theta^2 + \cos^2 \theta\, (\rmd \phi+i\dd\tau)^2 + \sin^2\theta \,\rmd \Omega_3^2\Big)\,,
	\end{equation}
	where $\rmd \Omega_3^2$ and $\rmd \tilde\Omega_3^2$ denotes the metrics on two copies of the round $S^3$.
Due to this 1/2-BPS twist, which locally is just a simple coordinate transformation but cannot be undone due to global restrictions on the Killing spinor, the four-form now takes a slightly altered form
	\begin{equation}
		C_4 = \rmi L^4 \sinh^4 \rho \,\rmd \tau \wedge \text{vol}_{\tilde S^3} + L^4\sin^4 \theta\, (\rmd \phi+ i\dd\tau)  \wedge {\text{vol}}_{S^3}\,.
	\end{equation}
It is worth pointing out here that the twist itself breaks supersymmetry by 1/2 as only half of the Killing spinors are independent of $\tau$ after performing the twist. The other half are projected out as Euclidean time is periodically identified.

If we now consider a probe D3-brane in this background, it is straightforward to see that a similar configuration of the D3-brane as above, i.e. wrapping $\tau$ and $\Omega_3$ with $\rho=0$, again solves the equations of motion.\footnote{There is another solution which wraps $\theta$, $\phi$ and the equator of $\Omega_3$. This solution has on-shell action $S_\text{cl} = N\beta$ so may contribute to the index as desired. This brane, however, does not preserve all the background supersymmetry. A $\kappa$-symmetry analysis, similar to the one preformed in Appendix \ref{App: kappa symmetry}, shows that such an embedding is $1/8$-BPS.} Furthermore, in Appendix \ref{App: kappa symmetry} we show that the embedding preserves all the supersymmetries that the background twist does as well. Comparing to the giant graviton solutions in Lorentzian signature \cite{McGreevy:2000cw,Grisaru:2000zn,Hashimoto:2000zp} the twist acts similarly to the velocity of the giant graviton. 

For the classical solution described we find an exact cancellation between the DBI and the WZ term in the D-brane action, and hence the on-shell action vanishes
\be
S_\text{cl} = 0\,.
\ee
This should already sound some warning bells as it indicates that the semi-classical expansion is problematic. One way to evade the vanishing of the on-shell action is to add a pure gauge contribution to $C_4$ of the form
\begin{equation}
	C_4^{\text{(new)}} = C_4 + \rmd \tau \wedge \text{vol}_{S^3}\,,
\end{equation}
in which case the on-shell action evaluates to
\begin{equation}
S_\text{cl}^{(\text{new})} = N \beta\,,
\end{equation}
as required from the QFT prediction of the 1/2-BPS index at finite $N$. However, one cannot argue for this addition in string theory merely based on the local symmetries of the background, instead one would have to study the global structure of the symmetry group preserved by the twisted background to show that the pure gauge transformation is correct. Despite this subtlety regarding the on-shell action we can carry on and compute the spectrum of fluctuations as we did for the Schur case above. The computation is analogous, and we skip all computational details and summarize our results for the spectrum in table \ref{Tab:halfBPS}.
	\begin{table}[h]
	\centering
	\begin{tabular}{lcccc}\hline\hline
	 4D Field& Degeneracy & $|Q|$ &  $M^2$ & KK mass  \\\hline
	 Scalars& $4$ & $0$   & $1$ & $n_\beta  $\\
	 & $2$ & $1$  & $1$ & $(n_\beta + \rmi  )$\\\hline
	 Fermions& $4(8)$ &  $1/2$  & $0$ & $(n_\beta + \rmi/2 )$\\ \hline
	 Vector& $1$ &  $0$  & $0$ &$n_\beta $\\
	\end{tabular}
	\caption{A summary of the spectrum of conformally coupled scalars, fermions, and vector on the D3-brane in the 1/2-BPS twisted background \eqref{halfBPSmetric}.}\label{Tab:halfBPS} 
	\end{table}

Using this spectrum, we can compute the one-loop partition function in the same way as we did above. We find that the Casimir energy is the same as for the Schur limit $E_c = -1$, but the single particle partition function is $\hat Z_1 = 1$. This simple result indicates that all non-zero modes exactly cancel leaving only the zero-mode contribution of two scalar fields. This would then indicate that the one-loop contribution of this brane scales as
\be
\mathcal Z_\text{D3}(\beta) \sim \rme^{-\beta E_c} N \,,
\ee
which does not match the leading non-perturbative term in \eqref{halfBPSIndex} which is $-q^{N+1}/(1-q)$.
Indeed since the on-shell action of the D3 vanishes, it is not clear whether this result is entirely trustworthy. The mismatch with the QFT is perhaps due to this issue.

Recently quantum D3-branes were analyzed in order to reproduce the 1/2-BPS index of ${\cal N}=4$ SYM in \cite{Eleftheriou:2023jxr}. There are a few differences when compared to our approach. Instead of doing the one-loop quantization of D3-branes as we do here, they employ supersymmetric localization to study the full D3-brane index. To this end they supersymmetrize the scalar zero-mode sector on the D3-branes using the expected supercharges preserved by the D3-brane. It is not clear to us whether this supersymmetrization gives a result that is consistent with the expansion of the fermionic D3-brane action. Upon a naive  comparison of their supersymmetric Lagrangian with our result in table \ref{Tab:halfBPS}, we were unable to match precisely the two theories, but since they only retain a subset of modes the comparison may be more involved than naively expected. Determining exactly what lies behind this mismatch is beyond the scope of this work. It is however encouraging to see that in \cite{Eleftheriou:2023jxr}, they are able to find a match with the QFT prediction.

	\section{D3-branes in the thermal Leigh-Strassler background}\label{Sec:LS}
	The second type IIB string theory setup in which we study Euclidean D3-branes is dual to the Leigh-Strassler CFT \cite{Leigh:1995ep}, the IR fixed point of an ${\cal N}=1$ RG flow originating from the $\mathcal N=4$ theory in the UV by turning on the mass for one chiral multiplet. The IR CFT preserves a $\U(1)$ R- and $\SU(2)$ flavor global symmetry. The bulk geometry is similar to the one discussed above, with, however, a squashed five-sphere and a reduced amount of preserved supersymmetry, and was first constructed in \cite{Khavaev:1998fb,Pilch:2000ej}. Explicitly, the metric given by
	\begin{equation}
		\rmd s^2 = L^2 \sqrt{w} \Big(\rmd s_{\text{AdS}_5}^2 + \frac23 \rmd s_{5}^2\Big )\,,
	\end{equation}
	where $\rmd s_{\text{AdS}_5}^2$ is identical to the AdS$_{5}$ metric in \eqref{Eq: 10d metric}, and thus we have imposed the same Schur twist as before to ensure that the background is supersymmetric.\footnote{It will be interesting in the future to study also the 1/2-BPS twist background for the Leigh-Strassler CFT.} Furthermore, we defined $w = 1+\sin^2\theta$, and the metric on the five-sphere is given by
	\begin{equation}
	\begin{aligned}
		\rmd s_5^2 = \rmd \theta^2 + \frac{\cos^2\theta}{2 w} \Big( \rmd\omega_1^2 + \sin^2\omega_1 \rmd \omega_2^2 + \frac{4w}{5w-2} (\rmd \omega_3 + \cos\omega_1 \rmd \omega_2)^2\\
		 + \frac{\sin^2 2\theta}{6 w(5w-2)} (\rmd \omega_3 + \cos\omega_1 \rmd \omega_2 + \frac{5w-2}{\cos^2\theta} \rmd\phi)^2 \Big)\,.
	\end{aligned}
	\end{equation}
	This metric reproduces the global symmetries in the dual field theory through isometries on the internal five-sphere. In particular, the $\SU(2)\times \U(1)$ symmetry is realized through a squashed $S^3\subset S^5$ spanned by the angles $\omega_i$. For future reference we also provide the chosen frame fields
	\begin{equation}
		\begin{split}
	\sum_{a=1}^5(e^a)^2&=L^2w^{1/2}\dd s_{\text{AdS}_5}^2\,,\quad e^6= Lw^{1/4} \sqrt{\frac{2}{3}}  \dd\theta\,,\\
	e^7&= \frac{1}{\sqrt{3}}Lw^{-1/4}  \cos\theta\, \dd \omega_1\,,\quad  e^8 = \frac{1}{\sqrt{3}}Lw^{-1/4}  \cos\theta\, \sin\omega_1\, \dd \omega_2\,,\\
	e^9&=  \frac{2}{\sqrt{3(5w-2)}}Lw^{1/4}   \cos\theta\, (\rmd \omega_3 + \cos\omega_1 \rmd \omega_2)\,,\\
	 e^{10}&=\frac{Lw^{1/4}}{3}\sqrt{2(5w-2)} \sin\theta \, \Big( \frac{\rmd \phi}{w} + \frac{\cos^2\theta}{w(5w-2)} (\rmd \omega_3 + \cos\omega_1 \rmd \omega_2)\Big)\,.
	\end{split}
	\end{equation}
	The axion vanishes and dilaton is constant in this background. To write down the RR and NSNS two-forms in a compact fashion it is useful to define the following complex two-form
	\begin{equation}
	\begin{aligned}
		B_2 + i C_2 = \frac{-L^2}{6}\sqrt{\frac{2}{3}} \rme^{-\rmi (\phi+\omega_3)}&\cos\theta (\rmd\omega_1 - \rmi \sin\omega_1 \rmd \omega_2) \\
		&\wedge \Big(2\rmd \theta + \frac{\rmi\sin2\theta}{w} (\rmd \phi - \rmd \omega_3 - \cos\omega_1 \rmd \omega_2)\Big)\,,
	\end{aligned}
	\end{equation}
	and finally, the four-form equals
	\begin{equation}
		C_4 = \frac{32 L^4}{27} \frac{\cos^4 \theta}{8 w} \sin\omega_1 \rmd \omega_1 \wedge \rmd \omega_2 \wedge \rmd \omega_3 \wedge \phi\,.
	\end{equation}
	The AdS length here is related to the rank of the gauge group and the string length through
	\begin{equation}
		L^4 = \frac{27 N \pi \ell_s^4}{8}\,.
	\end{equation}
	As mentioned in the introduction, it is expected that when computing the superconformal index of the Leigh-Strassler theory in the large $N$ limit that there are non-perturbative contributions which in the string theory description come from D3-branes wrapping the Euclidean time circle, and a three-cycle in the internal geometry, very much similar to the index of $\mathcal N=4$ SYM. The obvious choice for this three-cycle is the squashed three-sphere spanned by $\omega_i$. Extremizing the D3-brane action in this particular embedding shows that it localizes at $(\rho,\theta) = (0,0)$, where the worldvolume metric reduces to
	\begin{equation}\label{LSmet}
		\rmd s_{\text{D3}}^2 = L^2 \Big(\rmd \tau^2 + \frac13 (\rmd\omega_1^2 + \sin^2\omega_1\rmd\omega_2^2 + \frac43 (\rmd \omega_3 + \cos\omega_1 \rmd \omega_2)^2)\Big)\,.
	\end{equation}
	As in the previous case of standard AdS$_5\times S^5$, the D3-brane has the topology of $S_\beta^1 \times S^3$. Now however, the $S^3$ is squashed while preserving an $\SU(2) \times \U(1)$ isometry. The on-shell action of the brane is computed to
	\begin{equation}
		S_{\text{cl}} = \frac{3N \beta}{2}\,.
	\end{equation}
	In the dual field theory this result predicts the leading non-perturbative correction to the superconformal index to scale as
	\begin{equation}
		Z_{1\text{-loop}}\rme^{-3N\beta/2}\,,
	\end{equation}
	where $Z_{1\text{-loop}}$ is to be determined through a one-loop partition function of the fields living on the squashed D3-brane. We will take initial steps in determining this one-loop partition function by  first computing the spectrum. We apply the same  procedure and formulae as discussed in the previous section.

	In particular the  scalar fluctuations are determined using eqs. \eqref{Eq: pullbacks} and \eqref{bosonicaction}. The  Lagrangian is just a sum of Gaussian fields that are subject to the operator
	\begin{equation}\label{LSscalaroperator}
	{\cal K} = - D^2 + M^2\,,\qquad D_\mu = \nabla_\mu - Q i {\cal  A}_\mu - \tilde Q i \tilde {\cal  A}_\mu\,. 
	\end{equation}
	We now find two gauge fields, the first of them is a constant which is due to the Schur twist along AdS$_5$ while the second one is due to the non-trivial fibration in $S^5$ 
	\be
{\cal  A} = i\dd\tau\,,\qquad \tilde {\cal  A} = \f13 (\dd\omega_3+\cos\omega_1 \dd \omega_2)\,.
	\ee
	For each of the six scalar fields we must therefore specify their four-dimensional mass squared as well as the two charges $Q$ and $\tilde Q$. These are listed in table \ref{LSspectrum}.
	\begin{table}[h]
	\centering
	\begin{tabular}{lcccc}\hline\hline
	 4D Field& Degeneracy & $|Q|$ & $|\tilde Q|$ & $M^2$ \\\hline
	 Scalars& $2$ 		  & $0$   & $0$ & $1$\\
	 		& $2$ 		  & $1$   & $0$ & $1$\\
	 		& $2$ 		  & $3$   & $1$ & $2$\\\hline
	 Fermions& $2$ 		  & $3/2$   & $1/2$ & $0$\\
	 		 & $2$ 		  & $1/2$   & $1/2$ & $0$\\
	 \hline
	 Vector& $1$ &  $0$  & $0$ & $0$ \\
	\end{tabular}
	\caption{The spectrum of fluctuations around the classical D3-brane solution \eqref{LSmet} in the LS geometry.}\label{LSspectrum} 
	\end{table}
	For  the vector, unsurprisingly we find once more it to be massless with the standard Maxwell kinetic term as in \eqref{Eq: Maxwell term}. 

	To determine the fermionic spectrum it turns out to be convenient to choose a different  $\kappa$-symmetry gauge than discussed in previous section. This is because the three-form fields $H$ and $F_3$ contribute non-trivially to various terms in the fermionic action. However by choosing an appropriate kappa-symmetry gauge, all of these terms vanish and we are only left with the contribution from $F_5$ and the spin connection. This leaves us with massless fermions which like the scalar fields are charged with respect to the two gauge fields ${\cal A}$ and $\tilde{\cal A}$
	\begin{equation}
	i{\cal D} = i\slashed{D}\,,
	\end{equation}
	where $D$ is the same differential operator as appears for the scalar fields in \eqref{LSscalaroperator}. All fermions are charged with respect to the same gauge fields as the bosons and they are listed in table \ref{LSspectrum}.

	Using this spectrum, and the eigenvalues of the Laplacian on the $\SU(2)\times \U(1)$ squashed three-sphere (see for example \cite{Dowker:1998pi}) one can compute the full 1-loop determinant of the Euclidean D3-brane, providing a full prediction of the leading non-perturbative correction to the Leigh-Strassler Witten index, a computation we hope to report on in the near future.
	
	We would like to bring attention to the fact that the squashing of the three-sphere on which the D3-branes are wrapped preserves $\SU(2)\times\U(1)$ invariance in three-dimensions. Reducing our system to three dimensions employing a KK ansatz as we did above means that we should compute a 3D partition function on the squashed sphere. Such partition functions were  studied in \cite{Hama:2011ea} for superconformal field theories. In particular, it was shown that $\SU(2)\times\U(1)$ invariant squashing is Q-exact with respect to the localizing supercharge in three dimensions, which means that the full partition function only depends on the squashing parameter through a simple rescaling of length scales. It would be interesting to see if the partition function of the squashed D3-brane discussed in this section and the next section depend non-trivially on the squashing parameter or they show the `Q-exactness' reported in \cite{Hama:2011ea}. This would point to an underlying 3D supersymmetry of the D3-brane model which is not manifest in our formulation.

\section{D3-branes in S-fold backgrounds}\label{Sec: D3-branes in J-fold backgrounds}
The final example we will study is a non-geometric background of type IIB string theory. It is holographically dual to $\SU(N)$ $\mathcal N=4$ SYM compactified on a circle, with an S-duality twist. By compactifying the four-dimensional theory on the circle a 3D QFT is obtained that, in the IR, flows to an ${\cal N}=4$ CFT (which we refer to as the $\mathcal N=4$ S-fold), which can be understood as a gauging of the well-known Gaiotto-Witten theory in three dimensions \cite{Gaiotto:2008ak}. The three-sphere partition function of this theory was computed analytically as a function of $N$ in \cite{Assel:2018vtq,Ganor:2014pha}, and takes a form reminiscent of the Witten index of ${\cal N}=4$ SYM. Our goal in this section is to further study this $S^1_{\mathfrak{J}}\times S^3$ supersymmetric partition function of $\mathcal N=4$ SYM from a string theory perspective.\footnote{The subscript $\mathfrak J$ refers to the S-duality element used to twist the circle, whose explicit form we will provide below.}

The string theory background dual to the $\mathcal N=4$ S-fold was first constructed in \cite{Inverso:2016eet} and analyzed further in \cite{Assel:2018vtq}. Subsequently, in \cite{Bobev:2020fon,Bobev:2021yya,Arav:2021gra} it was shown that the 3D ${\cal N}=4$ SCFT lives on a ${\cal N}=2$ conformal manifold where at a generic point a $\U(1)_R \times \U(1)_F$ global symmetry is preserved. Apart from the supersymmetry enhanced $\mathcal N=4$ point, there is another symmetry enhanced point on the conformal manifold, where the flavor symmetry becomes $\text{SU}(2)_F$. In this section we will focus on this $\SU(2)_F\times \U(1)_R$ invariant point to study non-perturbative corrections to the three-sphere partition function. The field theory global symmetries are realized as isometries on the internal geometry in the bulk. In particular, the $\SU(2) \times \U(1)$ symmetry group is realized as a squashed $S^3 \subset S^5$, very much alike to the squashed three-sphere in the Leigh-Strassler background of the previous section. Exactly marginal operators are exact with respect to the supercharged used to localize three-dimensional theories on a sphere. Consequently the partition function is constant along the conformal manifold, and thus the answer in \eqref{JfoldPF} also holds in the $\text{SU}(2)$ enhanced point we study in this section. In \cite{Bobev:2023bxs} we showed that the non-perturbative corrections to the $S^3$ partition function scale as the on-shell action of D3-branes wrapping the compactified S-fold circle, and a three cycle inside $S^5$, much alike to the giant graviton expansion in the four-dimensional $\mathcal N=4$ SYM theory. In the following we will further analyze these D3-branes. 

We start with spelling out the full ten-dimensional background. The metric in Einstein frame takes the following form \cite{Bobev:2020fon,Guarino:2020gfe}
\be
\begin{split}
\dd s_{10}^2 &= L^2w^{1/4}\Big( \dd\varphi^2+\dd s_{\text{AdS}_4}^2 + \dd\theta^2 + \sin^2\theta\, \dd \phi^2+\cos^2\theta\,\dd \tilde\Omega_3^2\Big)\,,\\
\e^{\Phi} &= \e^{2\varphi}\f{1+\sin^2\phi\,\sin^2\theta}{\sqrt{w}}\,,\\
C_0 &= \f{\e^{-2\varphi}}{2}\f{\sin2\phi\,\sin^2\theta}{1+\sin^2\phi\,\sin^2\theta}\,,\\
C_2 &=  \e^{-\varphi} w^{-1/4}\Big[- \cos \phi\,\cos \theta \,  e^{7,10}-\sin \phi\,( \sin
   \theta e^{8,9}+e^{6,10})\Big]\,,\\
B_2 &=  \e^{\varphi} w^{-1/4}\Big[ - \sin \phi\,\cos \theta   e^{7,10} + \cos \phi\,(\sin \theta\, e^{8,9}+e^{6,10})\Big]\,,\\
C_4 &=\f{5-\cos2\theta}{4\sqrt{w}}\cot\theta \,e^{7,8,9,10}\,,
\end{split}
\ee
where, just as in the section before, we have defined $w = 1+\sin^2\theta$ and the squashed three-sphere metric equals
\be
\dd \tilde\Omega_3^2=  \f{1}{2w}(\dd \omega_1^2+\sin^2\omega_1\,\dd \omega_2^2 )+ \f14(\dd \omega_3+\cos\omega_1\,\dd\omega_2)^2\,,
\ee
is the metric on squashed $S^3$. This metric is round of unit radius when $w=2$. We have also defined the flat frame fields to simplify the expression for the form fields:
\be\begin{split}
e^1&= Lw^{1/8} \dd\varphi\,,\quad \sum_{a=2}^5(e^a)^2=L^2w^{1/4}\dd s_{\text{AdS}_4}^2\,,\quad e^6= Lw^{1/8}  \dd\theta\,,\\
e^7&= Lw^{1/8}  \sin\theta\, \dd \phi\,,\quad  e^8 = \f{Lw^{-3/8}}{\sqrt{2}}\cos\theta\,\dd\omega_1\,,\\
e^9&=  \f{Lw^{-3/8}}{\sqrt{2}}\cos\theta\,\sin\omega_1\,\dd\omega_2\,,\quad
 e^{10}= Lw^{1/8} \f{\cos\theta}{2}\, (\dd \omega_3+\cos\omega_1\,\dd\omega_2)\,.
\end{split}
\ee
Our coordinates take the following range
\be
\theta \in (0,\pi)\,,\qquad \omega_1\in (0,\pi)\,,\qquad \omega_2,\phi \in (0,2\pi)\,,\qquad \omega_3 \in (0,4\pi)\,.
\ee
Finally the AdS$_4$ length scale is set by the rank of the gauge group 
\be
L^4=2\pi N\ell_s^4\,.
\ee
The angle $\varphi$ is assumed to be periodic, with period $\beta$. It appears that the full background above does not allow for such periodic identifications since some of the fields have explicit dependence on $\varphi$. So in order to make the periodic identification we must simultaneously perform an $\SL(2,\R)$ rotation of the fields in order to ``glue'' the circle back onto itself. The $\SL(2,\R)$ transformation that is required takes the form \cite{Bobev:2020fon}
\be
\mathfrak{J} = \begin{bmatrix} \e^{-\beta}&0\\0&\e^\beta \end{bmatrix}\,.
\ee
In order for this transformation to be a valid transformation in string theory, it must take values in $\SL(2,\Z)$ which requires a different $\SL(2,\R)$ gauge for the background above. The end result is that
\be
\Tr (\mathfrak J) = 2\cosh\beta =k\,,
\ee
must be an integer and is mapped to a Chern-Simons level in the dual 3D conformal field theory. In this paper we will not bother changing to the proper string theory gauge which renders the $\mathfrak{J}$-matrix integer valued. This will not affect any of our results and if done, would only serve to complicate intermediate expressions.

Contrary to the pure AdS$_5$ solutions discussed in the previous two sections above, we do not need to perform any twist in order to preserve supersymmetry. Indeed in \cite{Bobev:2020fon} it was explicitly verified that the Killing spinor in the above background does not depend on $\varphi$ and so the compactification of that coordinate does not cause any issue for supersymmetry.

A non-trivial embedding of a D3-brane in this geometry was found in \cite{Bobev:2023bxs} which gives rise to non-perturbative corrections to the S-fold sphere partition function. The brane wraps $S^1_{\mathfrak{J}}$ as well as the squashed three-sphere $\dd\tilde\Omega_3^2$. It is easy to show that the embedding is stabilized at $\theta=0$ but the position of the D3-brane in AdS$_4$ is entirely unfixed. The brane world-volume metric takes the form\footnote{Once again we have stripped off the length scale $L$ in the D3-brane expressions and will reinstate them at the level of the tension.}
\be
\dd s_{D3}^2 =  \dd\varphi^2+ \f{1}{2}(\dd \omega_1^2+\sin^2\omega_1\,\dd \omega_2^2 )+ \f14(\dd \omega_3+\cos\omega_1\,\dd\omega_2)^2\,,
\ee
i.e. it is the metric on $S^1_{\mathfrak J}\times S^3$ where the metric on $S^3$ is squashed in such a way that an $\SU(2)\times \U(1)$ isometry is preserved, very much akin to the D3-branes wrapping Euclidean cycles in the Leigh-Strassler background. The on-shell action of the configuration is
\be
S_\text{cl} = N\beta\,.
\ee

Using the formulae in Sections \ref{ssec:bosons} and \ref{ssec:fermions} we can compute the spectrum of bosonic and fermionic fluctuations. Computing the bosonic spectrum is a simple application of equations (\ref{Eq: pullbacks},\ref{quadraticBosons}) and we obtain the scalar operators
\be
{\cal K} = - D^2 +M^2\,,\qquad D_\mu = \partial_\mu - Qi {\cal  A}_\mu\,,
\ee
and we find that some of our fields are charged with respect to a constant gauge potential which in this case takes the form:
\be\label{Jfoldgaugefield}
{\cal  A}= i\dd\varphi\,.
\ee
We find that four of the six scalar fields are massless and uncharged while the remaining two are charged with $Q=\pm 5/2$ and have mass  $M^2=9/4$.

Next consider the vector field. Largely the expansion of the vector field follows the same pattern as for AdS$_5$ except that now the rescaling of the vector fields to absorb $\e^{-\Phi/2}$ gives rise to a `connection' in the Maxwell Lagrangian. Ultimately this just leads to a shift of the KK modes $n_\beta\mapsto n_\beta+\rmi$ when we reduce the action to 3D.

When dealing with the fermions, it is convenient to return back to the action before fixing the kappa symmetry gauge \eqref{prekappafermions}. Once again the computation is more involved compared to the maximally supersymmetry thermal AdS background discussed in Section \ref{Sec: thermal AdS5} due to the presence on non-trivial three-forms field strengths $H$ and $F_3$ contributing to the fermion action. When combining the terms we find that the total contribution of three-forms to the action is\footnote{We gauge fix the classical value of $\phi$ to zero for simplicity.}
\be
L\e^{-\Phi/2}\slashed{H} \sigma_3 -L\e^{\Phi/2}\slashed{F}_3 \sigma_1 = \Gamma_{{\underline\varphi}{\underline\omega}_3}(\sigma_1 \Gamma_{\underline\phi}-\sigma_3 \Gamma_{\underline\theta} ) + 2\Gamma_{{\underline\omega}_1{\underline\omega}_2}(\sigma_1\Gamma_{\underline\phi} + \sigma_3 \Gamma_{\underline\theta})\,.
\ee
This term motivates that we use the operators $\sigma_1\Gamma_{\underline\phi}$ and $\sigma_3 \Gamma_{\underline\theta}$ to break up the ten-dimensional spinor when reducing to four dimensions. We use the two signs $s_{1,2}$ to denote the eigenvalues of the two operators respectively. Furthermore we must rescale the fermions by $\e^{\varphi/4}$ to absorb the explicit factor of $\e^{-\Phi/4}$ in the fermion action. This rescaling changes operator slightly since the $\varphi$-derivative acts on this factor. Doing so results in the fermionic operator
\be
i{\cal D} = \rmi \slashed{\nabla}_{S^1_J\times  S^3} - \f{3\rmi}4\Gamma_{\underline\varphi}\big(1-s_1 s_2\Gamma_{(4)}\big) + \f\rmi4\Gamma_{{\underline\varphi}{\underline\omega}_3}\Big((s_1-s_2)+2(s_1+s_2)\Gamma_{(4)}\Big)\,.
\ee
We note that the dirac operator on $S^1_{\mathfrak J}\times  S^3$ includes the spin connection on the squashed $S^3$. The terms that are proportional to $\Gamma_{\underline\varphi}$ can be interpreted as the coupling of the fermions to the constant gauge field \eqref{Jfoldgaugefield} whereas the other terms are a result of the squashing of the $S^3$. They give rise to mass terms when reduced down to the round two-dimensional sphere.

We summarize the end result of the bosonic and fermionic spectrum in Table \ref{Tab:Jspectrum}. We plan to report on the full 1-loop determinant of this effective theory on the squashed D3-brane in the near future. Similarly to the D3-branes in the thermal Leigh-Strassler background we find that the squashing of the three sphere is part of the class $\SU(2)\times\U(1)$ invariant squashed 3D partition functions studied in \cite{Hama:2011ea}. It is interesting to note that even though the D3-brane partition function should not depend on the position along the ${\cal N}=2$ S-fold conformal manifold \cite{Bobev:2021yya}, this seems like a highly non-trivial result in string theory. Even though the classical action of D3-branes has been verified to be independent of the position \cite{Bobev:2023bxs}, the D3-brane metric generically only exhibits $\U(1)\times\U(1)$ isometry. Moreover, we expect that the spectrum of fluctuations will depend non-trivially on the position along the conformal manifold. Nevertheless, the QFT prediction clearly indicates that the same answer for the D3-brane partition function must be obtained no matter the position. We may speculate that this invariance is a non-trivial consequence of some Q-exactness of the effective D3-brane theory or its 3D reduction in line with similar results in 3D localization \cite{Hama:2011ea}.
	\begin{table}[h]
	\centering
	\begin{tabular}{lccc}\hline\hline
	 4D Field& Degeneracy & $|Q|$ & $M^2$  \\\hline
	 Scalars& $4$ & $0$   & $0 $\\
	 & $2$ & $5/2$  & $9/4$\\\hline
	 Fermions& $1$ &  $3/2$  & $1/4$\\
	 		 & $1$ &  $3/2$  & $1$\\
			 & $1$ &  $0$    & $1/4$\\  
			 & $1$ &  $0$    & $1$\\  \hline
	 Vector& $1$ &  $1$  & $0 $\\
	\end{tabular}
	\caption{A summary of the spectrum of scalars, fermions, and vector on the D3-brane in the J-fold background.\label{Tab:Jspectrum} }
	\end{table}
\appendix
\section{Twisted supersymmetry and $\kappa$-symmetry for the giants}\label{App: kappa symmetry}
In this appendix we will study the supersymmetry preserved by the D3 giants in the 1/2-BPS and Schur-twisted backgrounds in Section \ref{Sec: D3s in thermal AdS}.\footnote{We thank Sameer Murthy and Mart\'i Rossell\'o for correspondence that motivated us to add the explicit $\kappa$-symmetry computation in this preprint.} For the twisted background discussed in \ref{Sec:LS} one could start from the results in \cite{Pilch:2004yg} to construct the Euclidean type IIB spinor that is preserved and do a similar analysis as we provide below. For the S-fold backgrounds discussed in Section \ref{Sec: D3-branes in J-fold backgrounds} the ten-dimensional spinor is currently not known, obstructing the $\kappa$-symmetry analysis of the D3-branes.\footnote{One way to construct the spinor in this background is to start from the spinor constructed in \cite{DHoker:2006vfr} and take a singular limit to find the analog spinor for the S-fold background, along the lines of the construction performed in \cite{Bobev:2020fon} for the bosonic fields of type IIB supergravity.}
\subsection{$1/2$-BPS twist}
We will start with the $1/2$-BPS giant described in Section \ref{Sec: 1/2 bps index}. To see that the twist preserves half of the supersymmetries one can employ a simple group theory argument. Namely, the supercharges of $\mathcal N=4$ SYM transform in the $({\mathbf 4},{\mathbf 4})$ representation of $\SO(2,4)\times\SU(4)$. We are interested in the theory compactified on $S^3\times S^1$. We must therefore consider the branching of ${\mathbf 4}=({\mathbf 2},{\mathbf 1})_1\oplus ({\mathbf 1},{\mathbf 2})_{-1}$ under $\text{SO}(4,2)\to \text{SO}(4)\times \text{SO}(2)\simeq \text{SU}(2)\times\text{SU}(2)\times\text{U}(1)$ where the $\text{U}(1)$ charges are indicated as subscripts. Performing the analogous breaking of the R-symmetry $\text{SU}(4)$ we find that the supercharges transform as
\begin{equation}
({\bf 2},{\bf 1},{\bf 2},{\bf 1})_{1,1}\oplus ({\bf 2},{\bf 1},{\bf 1},{\bf 2})_{1,-1}\oplus ({\bf 1},{\bf 2},{\bf 2},{\bf 1})_{-1,1}\oplus({\bf 1},{\bf 2},{\bf 1},{\bf 2})_{-1,-1}\,.
\end{equation}
it is clear from this that all supercharges are charged with respect to `time' and we cannot compactify it without breaking supersymmetry. Twisting together the two $\text{U}(1)$'s (i.e. time translations and the $\text{U}(1)$ R-symmetry) with magnitude $\pm1$ we can compactify the twisted time without breaking all supersymmetries. We only keep those that are uncharged with respect to the twisted time which are 8 out of 16 supercharges. We will now move on to explicitly construct the bulk spinors realising the breaking pattern mentioned above, and show that the D3-brane giant of interest preserves all supercharges that the background preserves.

For completeness we provide here the frame fields used to compute the Killing spinor
Finally, for completeness we specify the ten-dimensional frames $e^M$ and $\Gamma$-matrices that are useful when deriving the Killing spinor:
\begin{equation}
\begin{split}
e^1 &= L\,\dd\rho\,,\quad e^2 = L\sinh\rho\,\dd \xi_1\,,\quad e^3 = L\sinh\rho\,\sin\xi_1\,\dd \xi_2\,,\\
e^4 &= L\sinh\rho\,\sin\xi_1\,\sin\xi_2\,\dd \xi_3\,,\quad e^5 = L\cosh\rho\,\dd \tau\,,\\
e^6 &= L\,\dd \theta\,,\quad e^7 = L\sin\rho\,\dd \omega_1\,,\quad e^8 = L\sin\theta\,\sin\omega_1\,\dd \omega_2\,,\\
e^9 &= L\sin\theta\,\sin\omega_1\,\sin\omega_2\,\dd \omega_3\,,\quad e^{10} = L\cos\theta\, (\dd \phi + \rmi \rmd \tau)\,.
\end{split}
\end{equation}
And we parametrize the $\Gamma$-matrices as
\begin{equation}
	\Gamma_{m} = \sigma_1 \otimes \gamma_{m} \otimes \mathbf{1}_4\,,\quad \Gamma_{i} = \sigma_2 \otimes \mathbf{1}_4 \otimes \gamma_{i}\,, 
\end{equation}
with $m=1,\cdots,5$, $i=6,\cdots,10$, and the two sets of $\gamma$-matrices being the matrices on the respective five-dimensional spaces. In our conventions $\gamma_{12345}=\gamma_{6789\,10}=-\mathbf{1}_4$, and for later convenience we define
\begin{equation}
	\Gamma_{\text{AdS}_5} = \Gamma_{1\,2\,3\,4\,5}\,,\quad \text{and} \quad  \Gamma_{S^5} = \Gamma_{6\,7\,8\,9\,10}\,, \quad \Gamma_{11} = \rmi\Gamma_{\text{AdS}_5}\Gamma_{S^5}\,.
\end{equation}
With this explicit basis the background spinor that solves the IIB BPS equation
\begin{equation}
	\delta \psi_\mu =D_\mu \epsilon + \frac{\rmi}{16} \slashed{F}_5\,\sigma_2 \,\epsilon \,,
\end{equation}
where $D_\mu$ is the standard covariant derivative acting on spinors in ten dimensions, and the Killing spinor
\begin{equation}
	\epsilon = \left( \begin{aligned} \epsilon_1 \\ \epsilon_2 \end{aligned} \right)
\end{equation}
is a doublet spinors of positive chirality, with in total 32 independent real components. Explicitly in our background the spinor can be written as
\begin{equation}\label{KillingSpinor}
\begin{split}
\frac{1}{\sqrt{2}}(\epsilon_1 + \rmi \epsilon_2) &=  \rme^{\frac{-1}{2} \rho \Gamma_{\text{AdS}_5}\Gamma_1} \rme^{\frac{\rmi}{2} \theta \Gamma_{S^5}\Gamma_6}\rme^{\frac12 \xi_1 \Gamma_{1\,2}} \rme^{\frac12 \xi_2 \Gamma_{2\,3}} \rme^{\frac12 \xi_3 \Gamma_{3\,4}}   \rme^{\frac12 \omega_1 \Gamma_{6\,7}}  \\
&\qquad\qquad\qquad\qquad\cdot \rme^{\frac12 \omega_2 \Gamma_{7\,8}} \rme^{\frac12 \omega_3 \Gamma_{8\,9}} \rme^{\frac{\rmi}{2} \phi  \Gamma_{S^5} \Gamma_{10}}\rme^{\frac{-1}{2} \tau \Gamma_{\text{AdS}_5}\Gamma_{5}(1-\rmi  \Gamma_{5\,10})} \epsilon_0\,,
\end{split}
\end{equation}
where $\epsilon_0$ is a constant spinor of positive chirality in ten dimensions, i.e. a priori it has 32 independent real components. Since we have compactified the $\tau$-direction and want to preserve supersymmetry we must impose that spinor to be constant along this direction. In teh parametrisation provided above it is easy to see that this independence can be achieved by taking 
\begin{equation}\label{Eq: 1/2 bps twist projector}
	\frac12 (1 - \rmi \Gamma_{5\,10}) \epsilon_0 = 0\,,
\end{equation}
reducing the background supersymmetry by 1/2, making the Euclidean background 1/2-BPS.\footnote{An equivalent possibility to preserve 1/2-BPS is by using the "anti-twist" $\phi \rightarrow \phi - \rmi \tau$, which will impose the spinor to project onto $\mathcal P_{+} \epsilon_0 \equiv \frac12 (1 + \rmi \Gamma_{5\,10}) \epsilon_0 = 0$.} To study the supersymmetry preserved by the probe brane embedded as explained in Section \ref{Sec: 1/2 bps index}, we will use the following frames on the worldvolume of the brane
\begin{equation}
\hat e^1 = L\sqrt{\cosh^2\rho-\cos^2 \theta}\,\dd\tau\,,\quad \hat e^2 = e^7\,,\quad \hat e^3 = e^8\,,\quad \hat e^4 = e^9\,.
\end{equation}
The equations of motion on the brane fixes the worldvolume scalar $\rho = 0$, and imposes
\begin{equation}
	0 = 4 L^4 (1 - q_{\text{D3}}) \cos\theta \sin^3\theta\,,
\end{equation}
which has three classes of solutions:
\begin{equation}
	\text{I}: \quad q_{\text{D3}}=1\,,\quad \text{II}: \quad \theta = \pi/2 \,,\quad \text{III}: \quad \theta =0\,.
\end{equation}
We see that the metric of the brane degenerates for solution III, and thus we disregard it, while solutions I and II remain good. To inspect if the brane is supersymmetric one has to impose $\kappa$-symmetry
\begin{equation}
	\Gamma_{\text{D3}}\,\epsilon  = q_\text{D3} \epsilon\,,\quad \text{where} \quad  \Gamma_{\text{D3}} = \frac{1}{4!} \varepsilon^{abcd} \Gamma_{abcd}\, \sigma_2\,, 
\end{equation}
where we have denoted here the tangent directions on the brane with $a,b,\ldots$. Explicitly, the projector takes the form
\begin{equation}
\Gamma_{\text{D3}}= -\frac{1}{\sin\theta}\Gamma_{7\,8\,9}(\Gamma_5 + \rmi \cos\theta \Gamma_{10})\,\sigma_2\,,
\end{equation}
where we have fixed $\rho=0$ as dictated by the equations of motion.
%
The $\kappa$-symmetry projector has the same eigenspace as the twist projectors defined above and thus does not impose any additional constraints on the IIB spinor, making the brane 1/2-BPS, just as the background itself. The $\kappa$-symmetry constraint does however fix the charge of the brane to equal the background twist $q = 1$, making the on-shell action of the brane to vanish
\begin{equation}
	S_\text{cl} =0\,.
\end{equation}
The anti-brane, which has $q_{\text{D3}} = -1$, breaks supersymmetry completely, and the equations of motion will force the brane to sit at $\theta = \pi/2$, where the on-shell action of the anti-brane becomes
\begin{equation}
	S_{\text{cl}} = \frac{2\pi}{(2\pi \ell_s)^4} 4\pi^2 L^4 = 2N \beta\,.
\end{equation}
We end this section by noting that if one would have chosen to preserve the background supersymmetry with the anti-twist instead one would have found that the anti-brane preserves all supersymmetry, and the brane completely breaks supersymmetry.

\paragraph{Note:} We correct here a mistake written down in an older version of this preprint, where we had confused solution I and II together, taking at the same time $\theta=\pi/2$ and zero on-shell action. Supersymmetry however does not fix the value of the worldvolume scalar $\theta$. 
\subsection{Schur twist}
We can perform an analog analysis for the branes embedded in the Schur twisted background described in Section \ref{Sec: thermal AdS5}. The ten-dimensional frames in this case are taken to be
\begin{equation}
\begin{split}
e^1 &= L\,\dd\rho\,,\quad e^2 = L\sinh\rho\,\dd \xi_1\,,\quad e^3 = L\sinh\rho\,\sin\xi_1\,\dd \xi_2\,,\\
e^4 &= L\sinh\rho\,\sin\xi_1\,\sin\xi_2\, (\dd\xi_3 + \rmi \dd\tau )\,,\quad e^5 = L\cosh\rho\,\dd \tau\,,\\
e^6 &= L\,\dd \theta\,,\quad e^7 = L\sin\rho\,\dd \omega_1\,,\quad e^8 = L\sin\theta\,\sin\omega_1\,\dd \omega_2\,,\\
e^9 &= L\sin\theta\,\sin\omega_1\,\sin\omega_2\,\dd \omega_3 \,,\quad e^{10} = L\cos\theta\, \dd \phi\,.
\end{split}
\end{equation}
Using the same frames as before the spinor in this twisted background equals
\begin{equation}\label{KillingSpinor}
\begin{aligned}
	\frac{1}{\sqrt{2}}(\epsilon_1 + \rmi \epsilon_2)  =& \, \rme^{\frac{-1}{2} \rho \Gamma_{\text{AdS}_5}\Gamma_1} \rme^{\frac12 \xi_1 \Gamma_{1\,2}} \rme^{\frac12 \xi_2 \Gamma_{2\,3}} \rme^{\frac12\xi_3 \Gamma_{3\,4}}  \rme^{\frac{\rmi}{2} \theta \Gamma_{S^5}\Gamma_6} \rme^{\frac12 \omega_1 \Gamma_{6\,7}}  \\
	&\hspace{3cm} \cdot \rme^{\frac12 \omega_2 \Gamma_{7\,8}} \rme^{\frac12 \omega_3  \Gamma_{8\,9}} \rme^{\frac{\rmi}{2} \phi \Gamma_{S^5} \Gamma_{10}} \rme^{\frac{-1}{2} \tau \Gamma_{\text{AdS}_5} \Gamma_5(1+ \rmi \Gamma_{1\,2})} \epsilon_0\,.
\end{aligned}
\end{equation}
Just as before it is easy to see that the spinor becomes independent of the compactified $\tau$ circle if we take 
\begin{equation}\label{Eq: twist projector}
	\frac{1}{2}(1 + \rmi \Gamma_{1\,2}) \epsilon_0 = 0\,.
\end{equation}
The embedding of the brane that we are interested in for the Schur index has been outlined in the Section \ref{Sec: D3s in thermal AdS}. The frames we choose on the D3 brane in this case are
\begin{equation}
	\hat e^1 = L\sqrt{\cosh^2\rho - \sin^2 \xi_1 \sin^2\xi_2 \sinh^2 \rho}\,\,\dd\tau\,,\quad \hat e^2 = e^7\,,\quad \hat e^3 = e^8\,,\quad \hat e^4 = e^9\,.
\end{equation}
The equations of motion on the world volume of the brane force the scalars to take the one of the following two possible sets of values
\begin{equation}
	\text{I}:\quad  \rho = 0 \,,\quad \theta = \pi/2\,,\quad \text{II}: \quad \theta = \xi_1 = \xi_2 = \pi/2\,.
\end{equation}
The $\kappa$-symmetry analysis, however, shows that only configuration I (the configuration of interest in Section \ref{Sec: thermal AdS5}) is supersymmetric, in which case the $\kappa$-symmetry projector takes the form
\begin{equation}
	\Gamma_{\text{D3}} = \Gamma_{5\,7\,8\,9} \,\sigma_2\,,
\end{equation}
whose eigenspace differs from the projector in \eqref{Eq: twist projector}. In particular, the projector that solves the $\kappa$-symmetry equation is the same as we had found for the $1/2$-BPS twist in \eqref{Eq: 1/2 bps twist projector}. Subsequently, the D3-brane embedded in the Schur twisted background only preserves $1/4$ of the supersymmetries in type IIB. Just as was already stated in the case of the 1/2-BPS twist one can also perform an anti-twist, in which case the anti-D3-brane preserves 1/4 of the supersymmetries.

%

\bigskip
\newpage
\leftline{\bf Acknowledgements}
\smallskip
\noindent We are grateful to Francesco Benini, Nikolay Bobev, Giorgos Eleftheriou, Ohad Mamroud, Valentina Puletti, Sameer Murthy, and  Martí Rosselló for fruitful discussions. The research of FFG is supported by the Icelandic Research Fund under grant 228952-053 and is partially supported by grants from the University of Iceland Research Fund. JvM is supported by the ERC-CoG grant NP-QFT No. 864583 ``Non-perturbative dynamics of quantum fields: from new deconfined phases of matter to quantum black holes'' and by INFN Iniziativa Specifica ST\&FI. JvM also thanks the ITF at KU Leuven for hospitality during the preparation of this manuscript. FFG would like to thank the Isaac Newton Institute for Mathematical Sciences, Cambridge, for support and hospitality during the programme Black holes: bridges between number theory and holographic quantum information, where work on this paper was undertaken. This work was supported by EPSRC grant EP/R014604/1.

	\appendix
	\bibliography{GiantD3bib}
	\bibliographystyle{JHEP}
	
\end{document}